%% file: Sup_tex.tex
\documentclass{brad_cv}[2014/03/18]
\usepackage[unicode=true,
  linktocpage,
  linkbordercolor={0.5 0.5 1},
  citebordercolor={0.5 1 0.5},
  linkcolor=blue]{hyperref}
	
\input{macros}

\begin{document}


# Quasiparticle mass enhancement approaching optimal doping in a high-$T_c$ superconductor

B. J. Ramshaw[1*], S. E. Sebastian[2], R. D. McDonald[1], James Day[3], B. S. Tan[2], Z. Zhu[1], J.B. Betts[1], Ruixing Liang[3,4], D. A. Bonn[3,4], W. N. Hardy[3,4], N. Harrison[1]

Feb 23, 2015

[1]Mail Stop E536, Los Alamos National Labs, Los Alamos, NM 87545
[2]Cavendish Laboratory, Cambridge University, JJ Thomson Avenue, Cambridge CB3 OHE, U.K
[3]Department of Physics and Astronomy, University of British Columbia, Vancouver V6T 1Z4, Canada
[4]Canadian Institute for Advanced Research, Toronto M5G 1Z8, Canada
[*]Corresponding author: bradramshaw@gmail.com

In the quest for superconductors with higher transition temperatures ($T_c$), one emerging motif is that electronic interactions favourable for superconductivity can be enhanced by fluctuations of a broken-symmetry phase. Recent experiments have suggested the existence of the requisite broken symmetry phase in the high-$T_c$ cuprates, but the impact of such a phase on the ground-state electronic interactions has remained unclear. We use magnetic fields exceeding 90 tesla to access the underlying metallic state of the cuprate YBa$_2$Cu$_3$O$_{6+\delta}$ over a wide range of doping, and observe magnetic quantum oscillations that reveal a strong enhancement of the quasiparticle effective mass toward optimal doping. This mass enhancement results from increasing electronic interactions approaching optimal doping, and suggests a quantum-critical point at a hole doping of $p_{crit} \approx 0.18$.

In several classes of unconventional superconductors, such as the heavy fermions, organics, and iron pnictides, superconductivity has been linked to a quantum critical point (QCP). At a QCP, the system undergoes a phase transition and a change in symmetry at zero temperature; the associated quantum fluctuations enhance interactions, which can give rise to (or enhance) superconductivity [1, 2]. As the QCP is approached, these fluctuations produce stronger and stronger electronic correlations, resulting in an experimentally-observable enhancement of the electron effective mass [1, 3, 4, 5]. It is widely believed that spin fluctuations in the vicinity of an antiferromagnetic QCP are important for superconductivity in many heavy-fermion, organic, and pnictide superconductors [6, 2], leading to the ubiquitous phenomenon of a superconducting dome surrounding a QCP. The role of quantum-criticality in cuprate high-temperature superconductors is more controversial [7]: do the collapsing

experimental energy scales[8], enhanced superconducting properties (see Fig. 1), and evidence for a change in ground-state symmetry near optimal doping [9, 10, 11, 12, 13, 14, 15, 16] support the existence of strong fluctuations that are relevant to superconductivity [17, 18, 19, 2]? Alternative explanations for the phenomenology of the cuprate phase diagram focus on the physics of a lightly doped Mott insulator [7, 20], rather than a metal with competing broken-symmetry phases. Several investigations, both theoretical and experimental, suggest that competing order is present in the cuprates, and is associated with the charge (rather than spin) degree of freedom (such as charge density wave order, orbital current order, or nematicity, see Fig. 1) [12, 15, 16, 17, 18, 21, 22, 23, 24, 25, 26, 27, 28]. What has been missing is direct, low-temperature evidence that the disappearance of competing order near optimal doping, and the associated change in ground-state symmetry, is accompanied by enhanced electronic interactions in the ground state.

A powerful technique for measuring low-temperature Fermi surface properties is the magnetic quantum-oscillation phenomenon, which directly accesses quasiparticle interactions through the effective mass [29]. Such measurements have been successful in identifying mass enhancements near QCPs in lower-$T_c$ materials (e.g., CeRhIn$_5$ and Ba(FeAs$_x$P$_{1-x}$)$_2$ [3, 5]), but the robustness of superconductivity near optimal doping in the cuprates has impeded access to the metallic ground-state. The Fermi surface in underdoped cuprates is known to be relatively small and electron-like [30, 31, 32, 33, 34] in contrast to overdoped cuprates where a much larger hole-like surface is observed [35]. This suggests the existence of broken translational symmetry in the underdoped cuprates that "reconstructs" the large hole-like surface into the smaller

electron-like surface, and this translational symmetry breaking is likely related to the charge order observed in the same doping range as the small Fermi pockets[27, 15]. Thus it is desirable to perform a systematic study of the doping dependence of these small pockets as optimal doping is approached within a single cuprate family. We use high magnetic fields, extending to over 90 T, to suppress superconductivity and access quantum oscillations of the underlying Fermi surface over nearly twice the range of dopings than previously possible in a single family (Fig. 1).

We report the observation of quantum oscillations in YBa$_2$Cu$_3$O$_{6+\delta}$ at $\delta = 0.75$, $0.80$, and $0.86$ (hole doping $p = 0.135, 0.140,$ and $0.152$), shown in Fig. 2A. Three regimes are clearly seen in the data: zero resistance in the vortex-solid state; finite resistance that increases strongly with field in the crossover to the normal state; and, magnetoresistance accompanied by quantum oscillations in the normal state. We subtract a smooth and monotonic background from the magnetoresistance to obtain the oscillatory component [36]. One can make two immediate observations: (1) at higher doping the oscillation amplitude grows faster with decreasing temperature; and (2), the oscillation frequency changes very little between $p = 0.135$ and $p = 0.152$. The first observation directly indicates an increasing effective mass; the second observation constrains the doping where the reconstruction from large to small Fermi surface takes place. We quantify these observations below.

The evolution of the cyclotron effective mass with doping, and how it relates to temperature dependence of the quantum oscillations, can be understood quantitatively within

the Lifshitz-Kosevich formalism, which has been used successfully to analyse oscillations in cuprates at lower hole doping [23, 32, 33, 34, 37]. The effective mass extracted from the quantum oscillation amplitude [36] is plotted as a function of doping in Fig. 3C, which reveals an increase in the mass by almost a factor of three from p = 0.116 to p = 0.152. Note that electron-phonon coupling is generally observed to decrease with increasing hole doping in the cuprates, ruling it out as the mechanism of mass enhancement [38], and suggesting instead that the mass enhancement comes from increased electron-electron interactions. The enhancement of the effective mass toward p ≈ 0.18 is consistent with the doping-dependent maxima observed in the upper critical field $H_{c2}$ [39], and in the jump in the specific heat (Δγ) at $T_c$ [40, 41], both of which are expected to be enhanced by the effective mass through the density of states (see Fig. 1). Maxima in thermodynamic quantities are typical of quantum-critical systems at their QCPs, having been observed in many heavy fermion systems [42] and an iron pnictide superconductor [5]. The fact that some physical quantities do not show an enhancement toward optimal doping in $YBa_2Cu_3O_{6+\delta}$, such as the superfluid density [43], may be related to the fact that different physical properties experience different renormalizations from interactions [44], or the fact that they are measured in the superconducting state where the gap may serve as a cut-off.

The quantum oscillation frequency $F$ gives the Fermi surface area through the Onsager relation [29], $A_k = \frac{2\pi e}{\hbar} F$, where $A_k$ is the Fermi surface area in momentum space perpendicular to the magnetic field. In contrast to the effective mass, which is enhanced by almost a factor of three, the Fermi surface area only evolves weakly toward optimal doping: Fig.

2B shows $F$ increasing by roughly 20% from $p \approx 0.09$ to $p \approx 0.152$. The observation of the small Fermi surface pocket up to $p \approx 0.152$ requires that the reconstruction of the Fermi surface also persists up to this doping, strongly suggesting that the reconstruction is related to the incommensurate charge order also observed in this doping range [15, 25, 26, 27]. The large increase in effective mass with no accompanying large change in Fermi surface area is reminiscent of what is seen on approach to QCPs in CeRh$_2$Si$_2$, CeRhIn$_5$, and BaFe$_2$(As$_{1-x}$P$_x$)$_2$ [42, 5].

The connection between the mass enhancement we observe in quantum oscillations and high-$T_c$ superconductivity is evident in Fig. 4, which shows successive $T_c$ curves in increasing magnetic field. By 30 T—the third-highest curve in Figure 4—superconductivity persists only in two small domes centred around $p \approx 0.08$ and $p \approx 0.18$; by 50 T only the region around $p = 0.18$ remains. This phase diagram of YBa$_2$Cu$_3$O$_{6+\delta}$ in high field, with $T_c$ first suppressed to zero around $p \approx 0.125$, closely resembles that of La$_{2-x}$Ba$_x$CuO$_4$ in zero field, where static charge and stripe order are observed [45]. To emphasize the enhancement of the effective mass, we plot $1/m^*$ on this phase diagram (including previous $m^*$ measurements at lower doping [46]). This shows a trend toward maximum mass enhancement at $p \approx 0.08$ and $p \approx 0.18$—the same dopings at which superconductivity is the most robust to applied magnetic fields. One possible scenario for YBa$_2$Cu$_3$O$_{6+\delta}$ is that critical fluctuations surrounding $p_{crit} \approx 0.08$ and $p_{crit} \approx 0.18$ provide two independent pairing mechanisms, analogous to the two superconducting domes in CeCu$_2$Si$_2$ that originate at antiferromagnetic and valence-transition QCP [47]. A second scenario is a single underlying pairing mechanism whose

strength varies smoothly with doping[7, 48], but where T$_c$ enhanced at $p_{crit} \approx 0.08$ and $p_{crit} \approx 0.18$ by an increased density of states and/or by quantum-critical dynamics.

Our observed mass increase establishes the enhancement of electronic interactions approaching $p_{crit} \approx 0.18$. It is natural to ask whether this enhancement is caused by a quantum-critical point at $p_{crit} \approx 0.18$, and, subsequently, what is the associated broken-symmetry phase. The hole doping $p_{crit} \approx 0.18$ represents the juncture of several doping-dependent phenomena associated with underdoped cuprates. First, $p \approx 0.19$ represents the collapse to zero of energy scales associated with the formation of the pseudogap which onsets at temperature $T^*$[8]. Second, the onset of an anomalous polar Kerr rotation and neutron spin flip scattering both terminate at $p \approx 0.18$ [12, 13], representing an unidentified form of broken symmetry (that persists inside the superconducting phase for the Kerr experiment). Third, in high magnetic fields, the sign-change of the Hall coefficient in $YBa_2Cu_3O_{6+\delta}$ from positive to negative, and the anomaly in the Hall coefficient in $Bi_2Sr_{0.51}La_{0.49}CuO_{6+\delta}$, occur near p ≈ 0.18 [11, 49], suggesting that Fermi surface reconstruction from electron-like to hole-like occurs at this doping. Finally, $p \approx 0.18$ represents the maximum extent of incommensurate CDW order reported in several different experiments [26, 27, 15]. Although the Fermi surface reconstruction is likely related to this CDW order, its short correlation length and the weak doping dependence of its onset temperature appear to be at odds with the standard picture of long range order collapsing to $T = 0$ at a QCP [50]. Two scenarios immediately present themselves. In the first scenario, the suppression of superconductivity by an applied magnetic field allows the CDW to transition to long-range

order, as suggested by X-ray, NMR, and pulsed-echo ultrasound experiments [25, 26, 51]. In this first scenario, we would be observing a field-revealed QCP. In the second scenario, CDW order is co-existent with another form of order that also terminates near $p_{crit} \approx 0.18$. Such a coexistence is suggested by multiple experimental results, including but not limited to Nernst anisotropy[22], polarized neutron scattering [12] the anomalous polar Kerr effect[13]. In this second scenario, the CDW reconstructs the Fermi surface and the other hidden form of order drives quantum criticality. Regardless of the specific mechanism, and regardless of whether $p_{crit} = 0.18$ is a QCP in the traditional sense, the observation of an enhanced effective mass coincident with the region of most robust superconductivity establishes the importance of competing broken symmetry for high-$T_c$ superconductivity.

## Acknowledgements



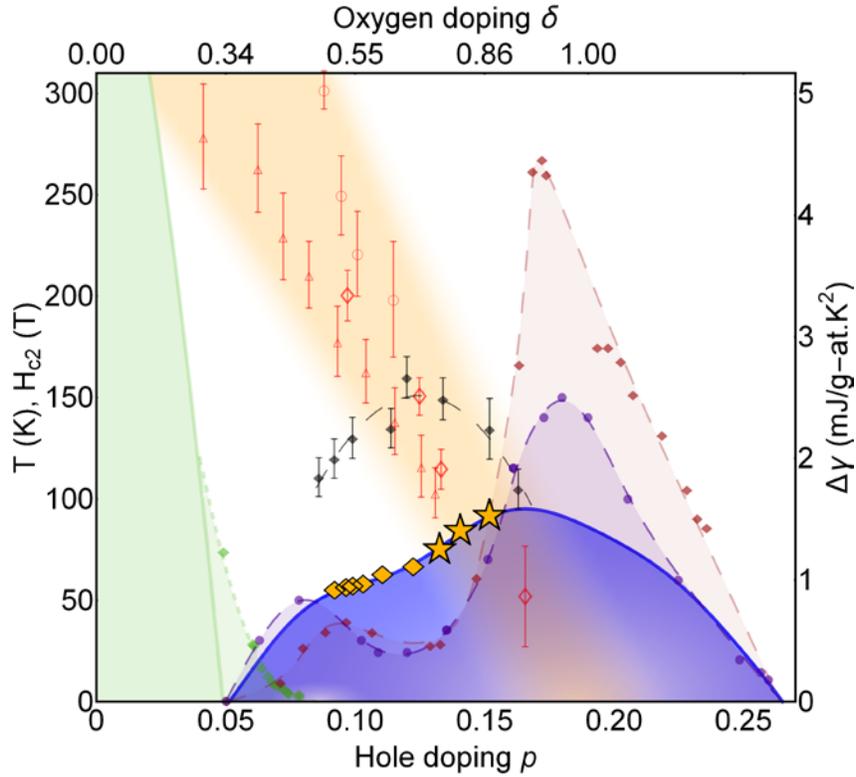

**Fig. 1: Cuprate temperature-doping phase diagram.** Long-range antiferromagnetic order (solid green line) gives way to superconductivity (solid blue line) near $p = 0.05$. Orange diamonds designate dopings where quantum oscillations have been observed previously[52, 53], and stars denote the new dopings presented in this paper. Short-range antiferromagnetic order (green diamonds) terminates at a quantum critical point at $p = 0.08$ [46, 54]; beyond $p = 0.08$, short-range charge order onsets above $T_c$ (solid black diamonds [15, 27]). The charge order, the onset of the pseudogap (as defined by neutron spin-flip scattering (open red circles)[12], the polar Kerr effect (open red diamonds[13]), and the change in the slope of resistivity with temperature (open red triangles[55])) terminate near $p = 0.18$, suggesting the possibility of a quantum critical point at this doping. Two thermodynamic quantities show enhancement near the critical dopings: the jump in the specific heat at $T_c$ ($\Delta\gamma$, maroon diamonds [40, 41]), and the upper critical field ($H_{c2}$, purple points [39]).

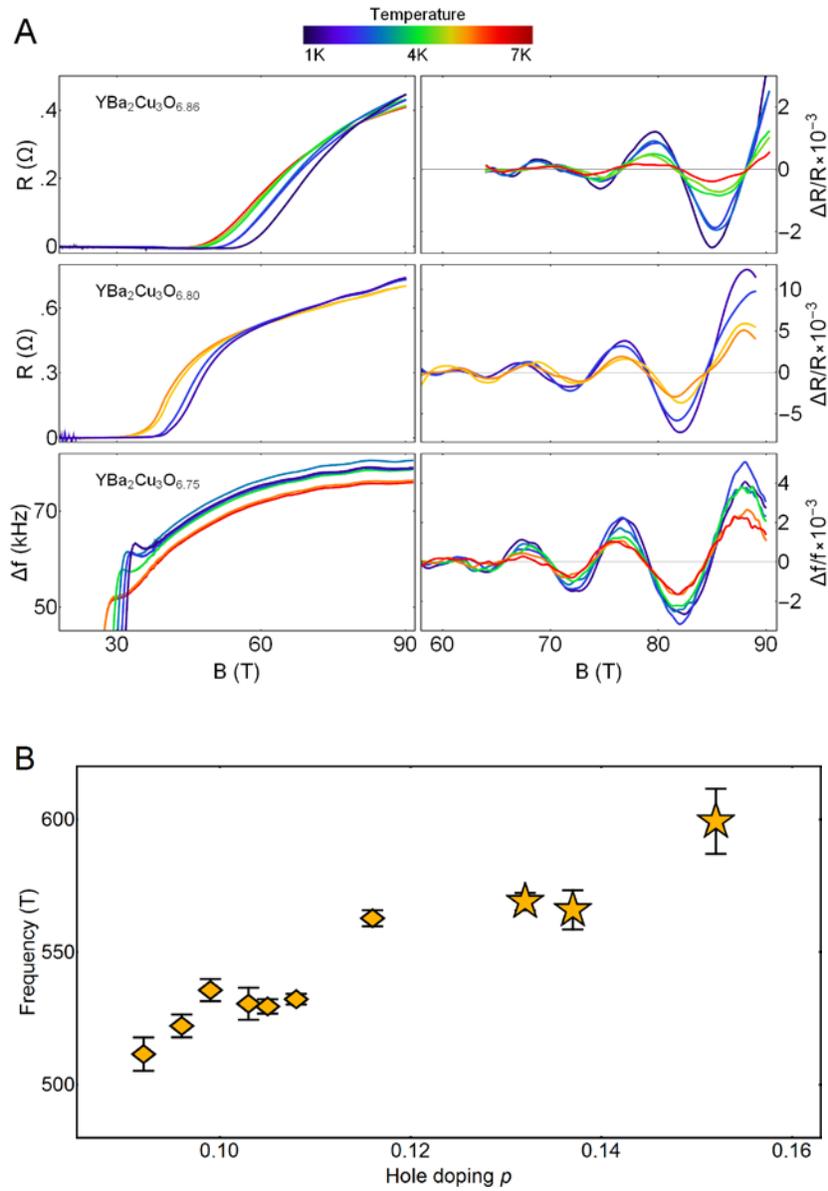

**Fig. 2: Quantum oscillations of the magnetoresistance in** $YBa_2Cu_3O_{6+\delta}$. (A) The bare (left panels) and oscillatory component (right panels) of the magnetoresistance. $\hat{c}$-axis transport was measured for $\delta = 0.80$ and $86$; skin depth, measured via frequency shift of an oscillatory circuit [36], was measured for $\delta = 0.75$. A smooth, non-oscillatory background is removed from the data to extract the oscillatory component [36]. The quantum oscillation amplitude is suppressed by a factor of two between 1.5 and 6 K in $YBa_2Cu_3O_{6.75}$, compared to a factor of five

over the same temperature range in YBa$_2$Cu$_3$O$_{6.86}$, indicating an increased effective mass for the higher doped sample. **(B)** Quantum oscillation frequency, proportional to Fermi-surface area, as a function of hole doping, with dopings below $p = 0.12$ taken from [56]. The frequencies and their uncertainties were obtained as described in the Materials and Methods [36].

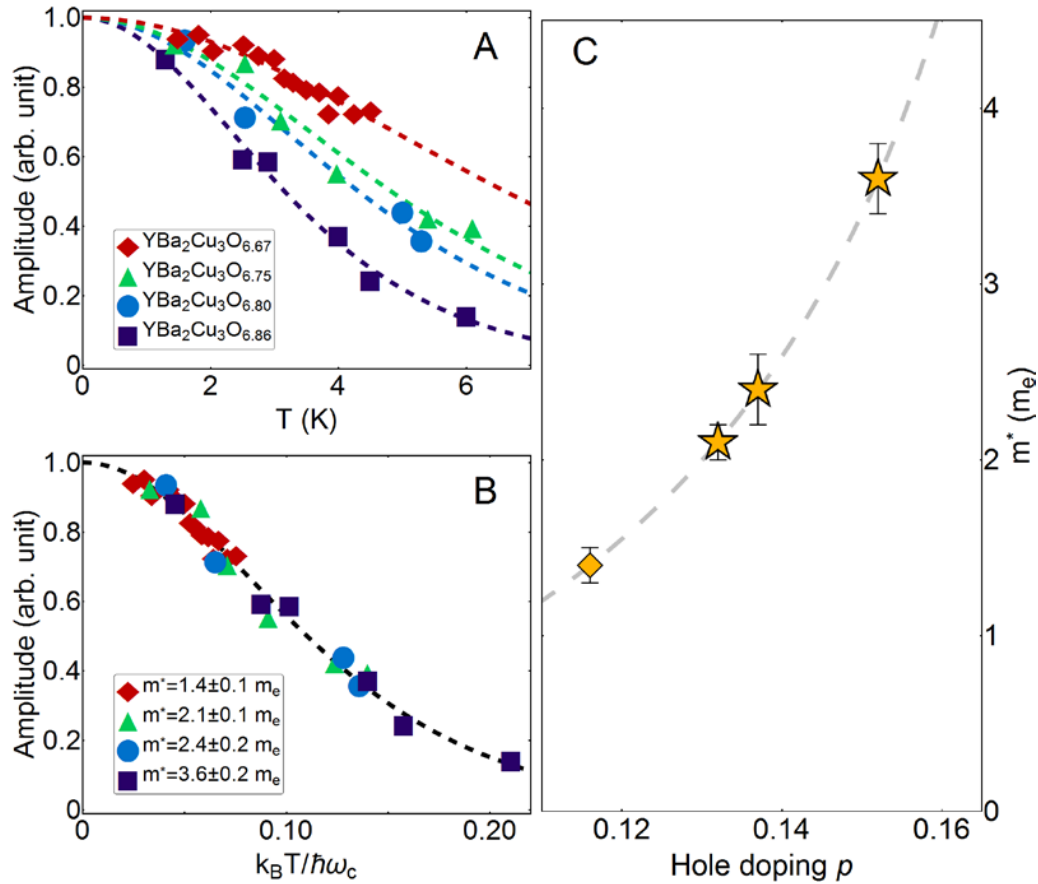

**Fig. 3: The quasiparticle effective mass in** YBa$_2$Cu$_3$O$_{6+\delta}$. Quantum oscillation amplitude **(A)** as a function of temperature, and **(B)** as a function of the ratio of thermal to cyclotron energy $k_B T/\hbar\omega_c$. Also included is detailed temperature dependence of YBa$_2$Cu$_3$O$_{6.67}$—a composition at which oscillations have previously been reported [53]. (A) illustrates the increase in $m^*$ with

increased hole doping, with fits to Eq. 1. (B) shows the same data versus $k_B T/\hbar\omega_c^*$, where $\omega_c^* = eB/m^*$: this scaling with $m^*$ shows the robustness of the fit across the entire doping and temperature range. **(C)** The effective mass as a function of hole doping; error bars are the standard error from regression of Eq. 1 to the data. The dashed line is a guide to the eye.

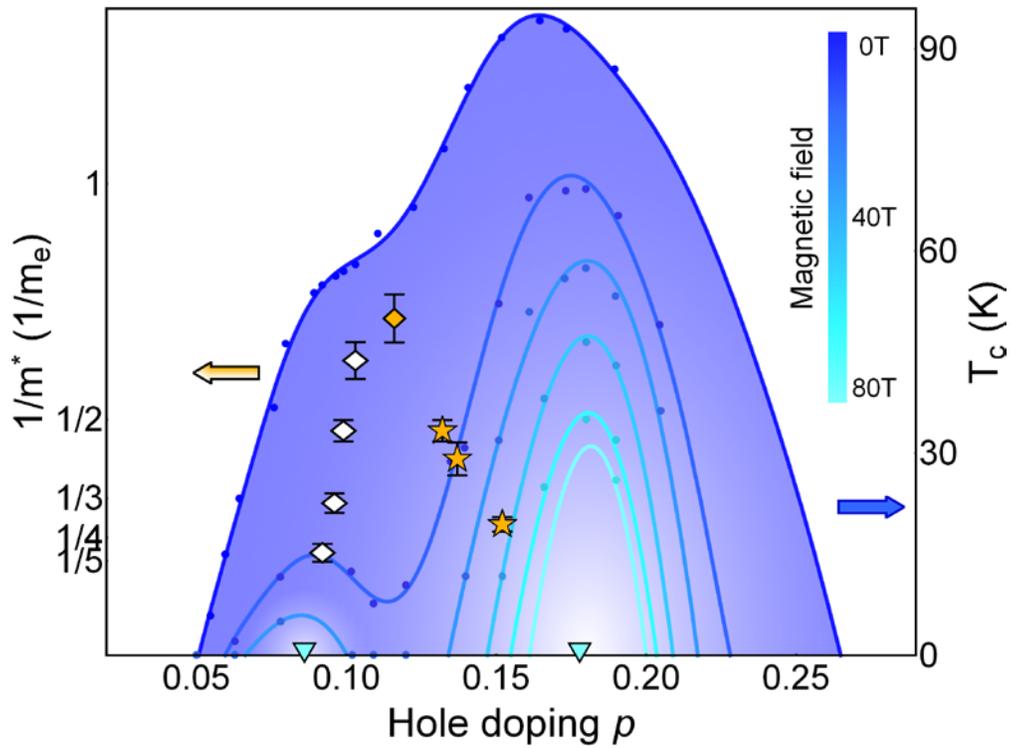

**Fig. 4: A quantum critical point near optimal doping.** The blue curves correspond to $T_c$, as defined by the resistive transition (right axis), at magnetic fields of 0, 15, 30, 50, 70, and 82 T (some data points taken from [39], [57].) As the magnetic field is increased, the superconducting $T_c$ is suppressed. By 30 T two separate domes remain, centred around $p \approx 0.08$ and $p \approx 0.18$; by 82 T only the dome at $p \approx 0.18$ remains. The inverse of the

effective mass has been overlaid on this phase diagram (left axis), extrapolating to maximum mass enhancement at at $p \approx 0.08$ and $p \approx 0.18$ (white points taken from [56]). This makes explicit the connection between effective mass enhancement and the robustness of superconductivity.

**Supplementary Materials**
    Materials and Methods
    Supplementary Text
    Figs. S1 to S15
    References [59-69]


\nocite{*}


\title{Supplementary Information: Quasiparticle mass enhancement approaching optimal doping in a high-\Tc superconductor}
\author{B.~J.~Ramshaw, S.~E.~Sebastian, R.~D.~McDonald, James Day, B.~Tan, Z.~Zhu, J.B.~Betts,\\ Ruixing Liang, D.~A.~Bonn, W.~N.~Hardy, N.~Harrison}
\maketitle

\section{Materials and Methods}
\label{se:Materials and Methods}
\subsection{Samples:} \YBCOd samples were prepared for transport measurements in the same manner as in our previous studies.\citep{Doiron-Leyraud:2007,Ramshaw:2011} Prior to measurement, samples were heated above the ortho-III and ortho-VIII phase transition temperatures\citep{Zimmermann:2003} to disorder the oxygen in the chains, and then quenched in liquid nitrogen. This process changes the nature of the oxygen defects in \YBCOd while preserving the total oxygen content, and enabled the observation of quantum oscillations by increasing the quasiparticle lifetime. 

\subsection{Measurement:} Four-point \caxis electrical resistance was measured for \YBCO{80} and \YBCO{86} using a digital lock-in amplifier. Skin-depth, proportional to $\hat{a}\!-\!\hat{b}\!-$plane resistance, was measured using a proximity detector oscillator on \YBCO{67} and \YBCO{75}.\citep{Sebastian:2010} All pulsed field measurements---up to 92 T for \YBCO{86}, \YBCO{80}, and \YBCO{75}; up to 65 T for \YBCO{67}---were made at the National High Magnetic Field Laboratory--Pulsed Field Facility. 

\subsection{Analysis:} The temperature-dependent amplitude $A\!\left(T\right)$ is extracted from the oscillatory magnetoresistance by fitting the standard Lifshitz-Kosevich expression for a quasi-2D Fermi surface\citep{Ramshaw:2011} $\frac{\Delta R}{R} = A\!\left(T\right)e^{-\frac{\pi}{\omega_c \tau}}\cos\!\left(\frac{2\pi F}{B}\right)J_0\!\left(\frac{2\pi \Delta F}{B}\right)$ at each temperature, keeping $\tau$, $F$, and $\Delta F$ fixed for a particular doping. The mass is then obtained by fitting the amplitude to $A\!\left(T\right) = \frac{2\pi^2 \frac{k_B T}{\hbar \omega_c^\star}}{\sinh\left(2 \pi^2\frac{k_B T}{\hbar \omega_c^\star}\right)}$.

 Because of the small number of oscillations available at high doping, the frequency $F$ was obtained in three different ways to check for consistency: by fitting the oscillatory component to the Lifshitz-Kosevich expression above; by Fourier-transforming the oscillatory data; and by Landau-indexing the oscillation peak positions \citep{Vignolle:2011b}. The uncertainties in $F$ shown in Fig. 2B were obtained from the Fourier transform peak widths. While there are systematic differences of about 3\% between the three methods of frequency determination, the trend of $F$ with hole doping $p$ is the same to within the uncertainty.

\section{Supplementary Text}
The Fermi surface of a layered material, such as a cuprate \highTc superconductor, is generally composed of cylinders that extend along the inter-layer direction, with finite interlayer tunnelling warping the cylinders \citep{Bergemann:2003}.  Quantum oscillations, which are sensitive to extremal Fermi surface areas perpendicular to an applied magnetic field, are used to map out this geometry in detail. The total magnetoresistance $R(B)$ of \YBCOd, with both field and current applied parallel to the \caxis, is composed of a background magnetoresistance, $R_0(B)$, with an oscillatory component $A_{osc} (B)$ that is scaled by the size of the background resistance\cite{Shoenberg:1984}:
\begin{equation}
R(B) = R_0(B)\left(1+A_{osc}(B)\right). 
\label{eq:1}
\end{equation}
The oscillatory component $A_{osc}(B)$ for a simple quasi-2D Fermi surface is described within the Lifshitz-Kosevich formalism as (see \citep{Ramshaw:2011}):
\begin{align}
A_{osc}(B) &= A R_T R_D \cos\!\left(\frac{2\pi F}{B}\right)J_0\!\left(\frac{2\pi \Delta F}{B}\right)  \label{eq:2}\\
R_D &= e^{-\frac{\pi}{\omega_c \tau}} \label{eq:3}\\
R_T &= \frac{2\pi^2 \frac{k_B T}{\hbar \omega_c}}{\sinh\left(2 \pi^2\frac{k_B T}{\hbar \omega_c}\right)} ,\label{eq:4}
\end{align}
where $A$ is an amplitude factor, $F$ is the oscillation frequency (proportional to the area of the Fermi surface perpendicular to the applied field),  $\Delta F$ is the first harmonic of warping in the $k_z$ direction,  $\omega_c\equiv e B/\mstar$ is the cyclotron frequency, \meff is the cyclotron effective mass, and $\tau$ is the quasiparticle lifetime.  Since we are only interested in the \textit{temperature} dependence of the oscillation amplitude, the exact Fermi surface geometry (which constrains the \textit{field} dependence) is relatively unimportant here \cite{Audouard:2009,Ramshaw:2011,Sebastian:2014}. Since the masses for different frequency components have been reported to be very similar \cite{Sebasian:2012}, the use of \autoref{eq:2} is preferable as a minimal model that prevents ``over-parametrization'' of the data set. 

The desired component in \autoref{eq:1} is the oscillatory component, $A_{osc}(B)$, and thus the background magnetoresistance $R_0(B)$ must be removed from the data. This is done by fitting \autoref{eq:1} to the measured data at each temperature, together with a polynomial (generally 3rd or 4th order) for the background:
\begin{equation}
R(B) = \left(a_0 + a_1 B + a_2 B^2 + a_3 B^3 ...\right) \left(1+A_{osc}(B)\right), 
\label{eq:5}
\end{equation}
where the coefficients $a_i$ are free parameters, and $A_{osc}(B)$ contains the Fermi surface parameters $A$, $F$, $\Delta F$, \meff, and $\tau$ (see \autoref{eq:1}). The oscillatory component $A_{osc}(B)$ is then obtained by dividing out the background polynomial and subtracting $1$ from the data. Because the fit parameters $A$, $F$, $\Delta F$, \meff, and $\tau$ are generally temperature-independent \cite{Shoenberg:1984}, only the polynomial coefficients $a_i$ vary as the background magnetoresistance changes with temperature.

\subsection{Data and fits}
The following four sections present all of the fits to the data for the dopings discussed in the main text, including new data taken for \YBCO{67} to precisely determine the mass (not shown in Figure 2a of the main text because of the lower field range). For each doping, the vertical span for the plots at each temperature is the same: this allows the evolution of the oscillation amplitude to be tracked by eye without background subtraction. For \YBCO{80} and \YBCO{86}, where the oscillation amplitude becomes small due to the increase in effective mass, we also present the derivatives of the data. 

\subsection{\YBCO{67}}
Quantum oscillations were measured in \YBCO{67} using the proximity detector oscillator  (PDO) technique. This is a contactless resistivity probe, where the sample is placed next to a pickup coil that is part of an oscillatory circuit. Changes in sample resistance as a function of magnetic field change the sample skin depth, tuning the inductance of the coil and producing a frequency shift in the oscillator self-resonance. 

\begin{figure}[H]
\includegraphics[width=\columnwidth]{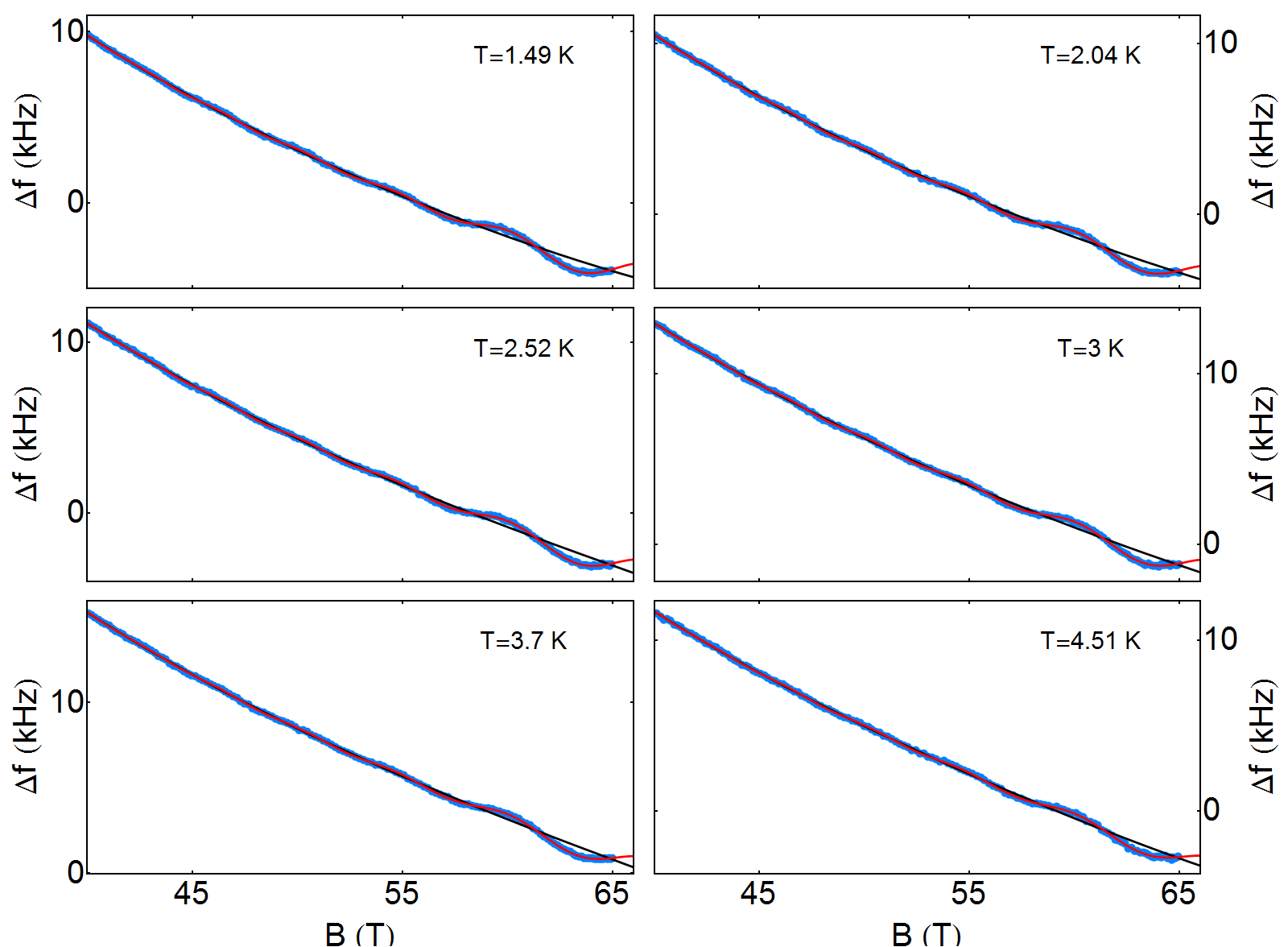}
\caption{ Shift in the proximity-diode oscillator frequency for \YBCO{67}, proportional to the skin depth (and therefore resistivity) at $\sim 30$ MHz: the raw data is in blue; the fit to \autoref{eq:2} plus a background is in red; the background alone is in black. The data at six temperatures, spanning the entire temperature range, are shown here. All panels have the same vertical span.}
\label{fig:67_1}
\end{figure}

\begin{figure}[H]
\includegraphics[width=\columnwidth,clip=true, trim = 0mm 60mm 0mm 60mm]{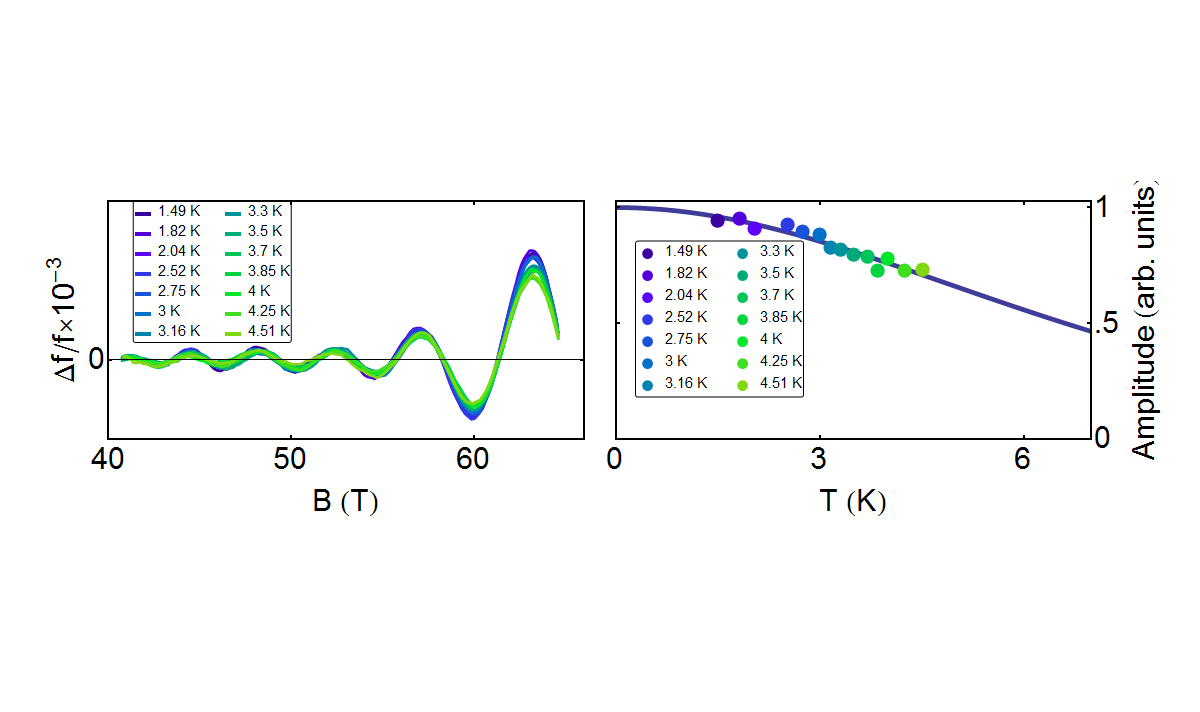}
\caption{ \textit{Left panel:} Relative frequency shift with respect to the background, proportional to $\Delta \rho / \rho$, for \YBCO{67} at each temperature with a background subtracted. \textit{Right panel:} Fit to the oscillation amplitude, yielding $\meff = 1.4\pm0.1$. }
\label{fig:67_2}
\end{figure}

\autoref{fig:67_1} shows the raw frequency-shift data for \YBCO{67} at six of the fourteen measured temperatures (spanning the full temperature range), along with the polynomial background in black and the full fit to \autoref{eq:5} in red. The fits yield $F=563 \pm 3$~T, $\Delta F = 15 \pm 6$~T, $\tau=0.08 \pm 0.01$~ps, and $\mstar = 1.4 \pm 0.1$~m$_e$.  Of the four dopings presented here, only \YBCO{67} has a significant contribution from a third frequency of $F = 519 \pm 5$ T. The background-free data from these fits at all temperatures, along with the mass fit from the oscillation amplitude, is shown in \autoref{fig:67_2}. 

\subsection{\YBCO{75}}
The same contactless magnetotransport technique that was performed on \YBCO{67} was also performed on \YBCO{75} (during a different experiment run and in a different magnet, hence the different temperatures and maximum field). 

\begin{figure}[H]
\includegraphics[width=\columnwidth]{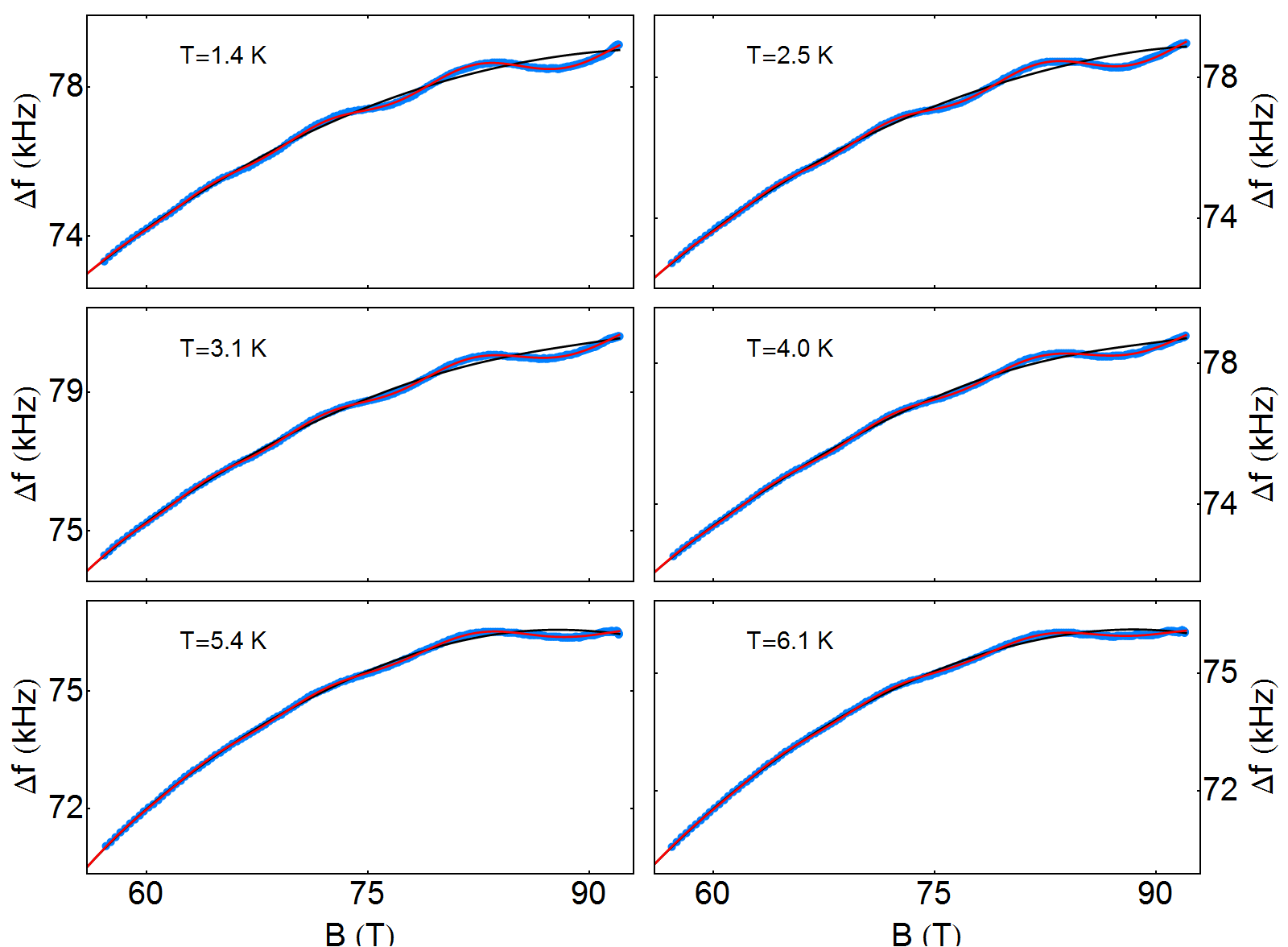}
\caption{Shift in the proximity-diode oscillator frequency for \YBCO{75}, proportional to the skin depth (and therefore resistivity) at $\sim 30$ MHz: the raw data is in blue; the fit to \autoref{eq:2} plus a background is in red; the background alone is in black. All panels have the same vertical span.}
\label{fig:1}
\end{figure}

\begin{figure}[H]
\includegraphics[width=\columnwidth,clip=true, trim = 0mm 60mm 0mm 60mm]{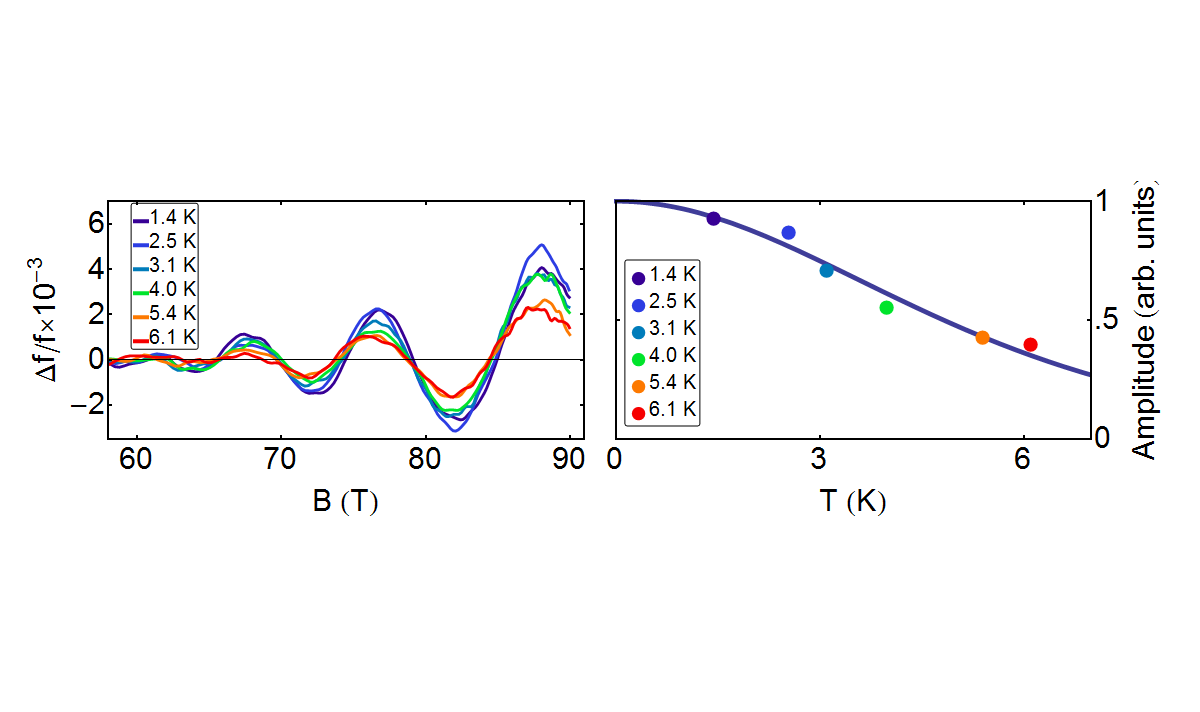}
\caption{ \textit{Left panel:} Relative frequency shift with respect to the background, proportional to $\Delta \rho / \rho$, \YBCO{75} at each temperature with a background subtracted. \textit{Right panel:} Fit to the oscillation amplitude, yielding $\meff = 2.1\pm0.1$. }
\label{fig:2}
\end{figure}

\autoref{fig:1} shows the raw frequency-shift data for \YBCO{75} at each temperature, along with the polynomial background in black and the full fit to \autoref{eq:5} in red. The fits yield $F=569 \pm 4$~T, $\Delta F = 19 \pm 7$~T, $\tau=0.06 \pm 0.02$~ps, and $\mstar = 2.1 \pm 0.1$~m$_e$. At 70 T these parameters yield $\wc \tau \approx 0.5$, consistent with oscillations becoming visible near this field.  While the shape of the background evolves with increasing temperature (due to the temperature dependence of the non-oscillatory magnetoresistance), it remains smooth and non-oscillatory at all temperatures. The background-free data from these fits, along with the mass fit from the oscillation amplitude, is shown in \autoref{fig:2}. 

\subsection{\YBCO{80}}
\begin{figure}[H]
\includegraphics[width=\columnwidth]{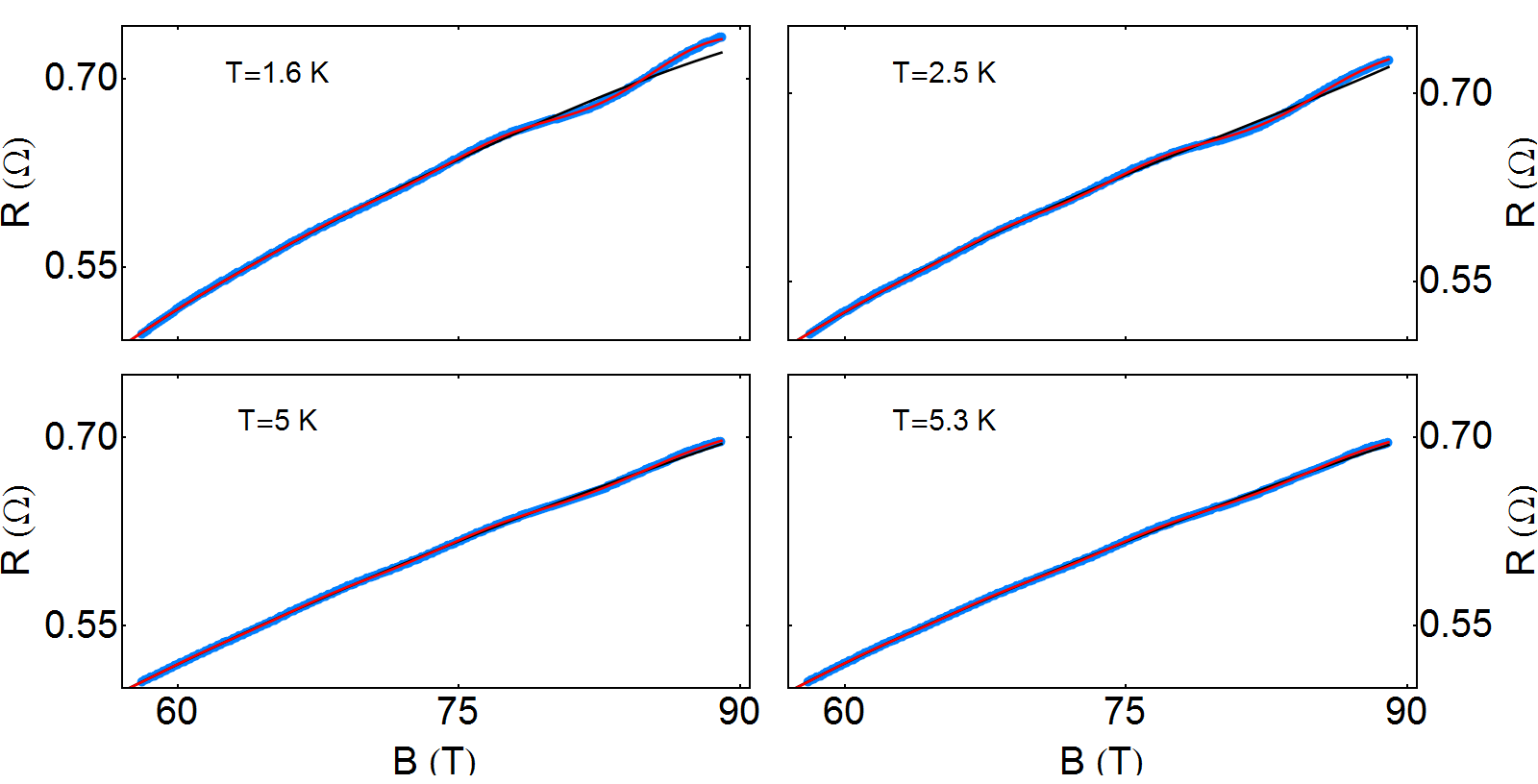}
\caption{ \caxis magnetotransport for \YBCO{80}: the raw data is in blue; the fit to \autoref{eq:2} plus a background is in red; the background alone is in black. All panels have the same vertical span.}
\label{fig:3}
\end{figure}

\begin{figure}[H]
\includegraphics[width=\columnwidth,clip=true, trim = 0mm 60mm 0mm 60mm]{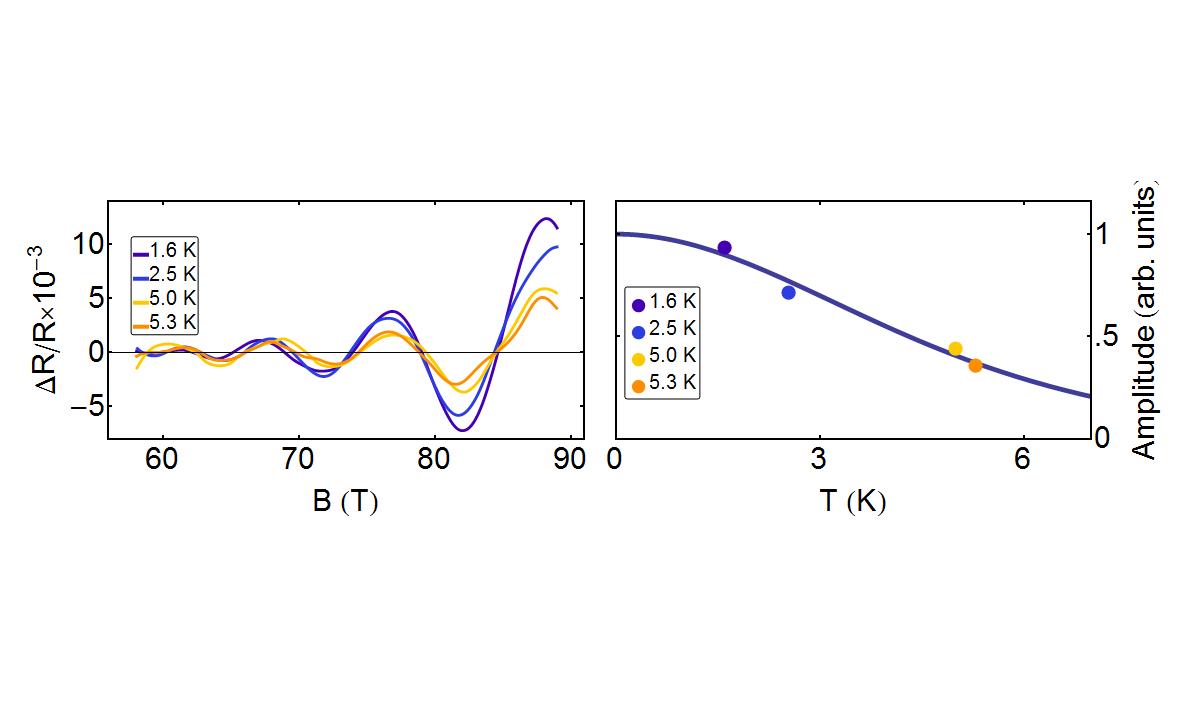}
\caption{ \textit{Left panel:} Magnetoresistance for \YBCO{80} at each temperature with a background subtracted. \textit{Right panel:} Fit to the oscillation amplitude, yielding $\meff = 2.4\pm0.2$. }
\label{fig:4}
\end{figure}

Shubnikov-de Haas oscillations were measured in the \caxis resistivity of \YBCO{80}. Sample contacts were prepared in a Corbino-like geometry, with large current contacts on both faces and small voltage contacts in the centre (see \citet{Ramshaw:2012} for sample preparation details).   Approximately 3 mA of current was driven through the sample at 500 kHz, and the voltage signal across the sample was recorded at 20 MHz using a digitizer. The data was then processed with a digital lock-in amplifier. 
\autoref{fig:3} shows the raw magnetoresistance data for \YBCO{80} at each temperature, plus the polynomial background in black and the full fit to \autoref{eq:5} in red. The fits yield $F=565 \pm 8$~T, $\Delta F = 21 \pm 8$~T, $\tau=0.09 \pm 0.02$~ps, and $\mstar = 2.4 \pm 0.2$~m$_e$. The data with the background removed, along with the mass fit, is shown in \autoref{fig:4}.

\begin{figure}[H]
\includegraphics[width=\columnwidth,]{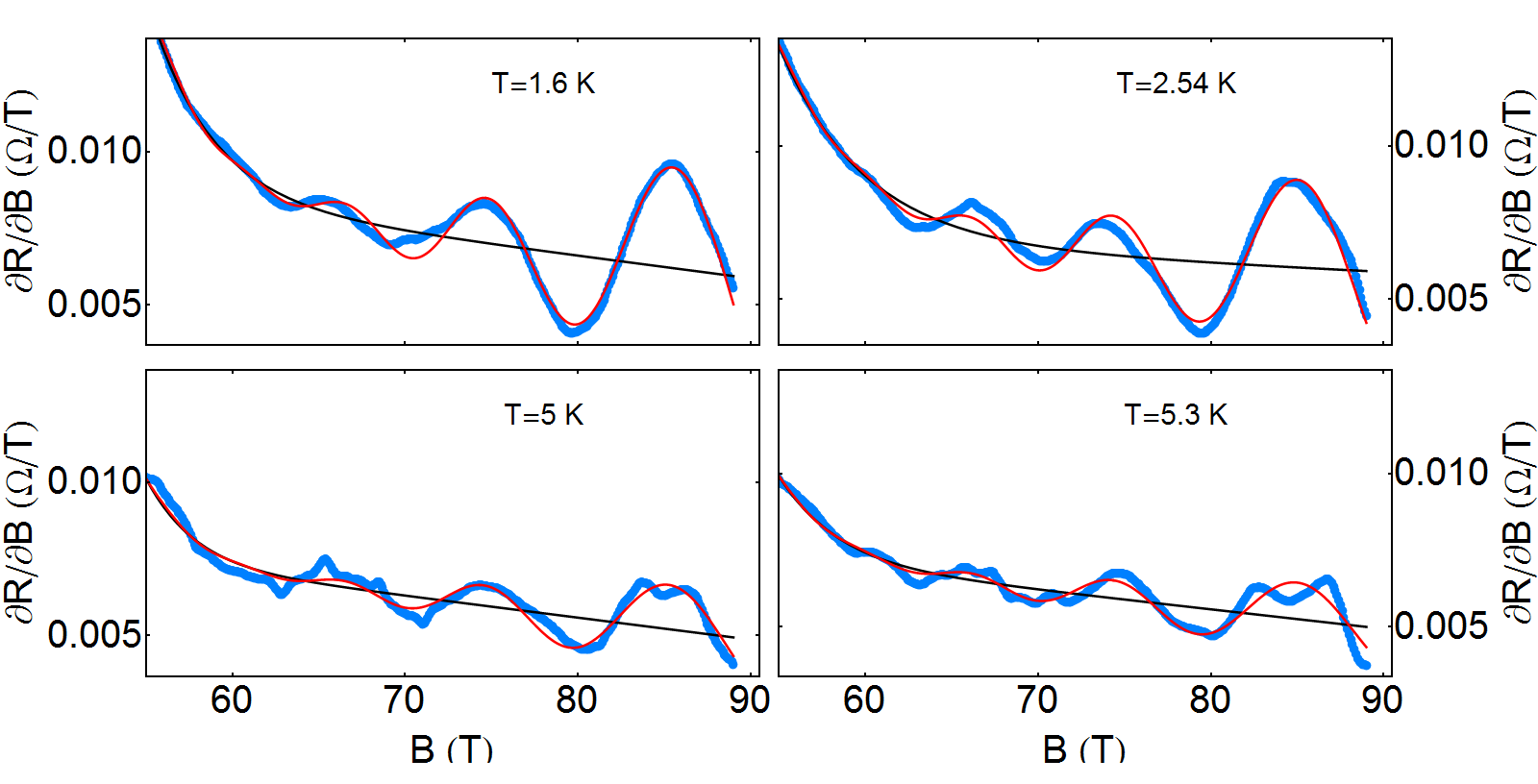}
\caption{Derivative of the magnetoresistance for \YBCO{80}: the raw data is in blue; the fit to \autoref{eq:2} plus a background is in red; the background alone is in black. All panels have the same vertical span.}
\label{fig:3b}
\end{figure}

\begin{figure}[H]
\includegraphics[width=\columnwidth,clip=true, trim = 0mm 60mm 0mm 60mm]{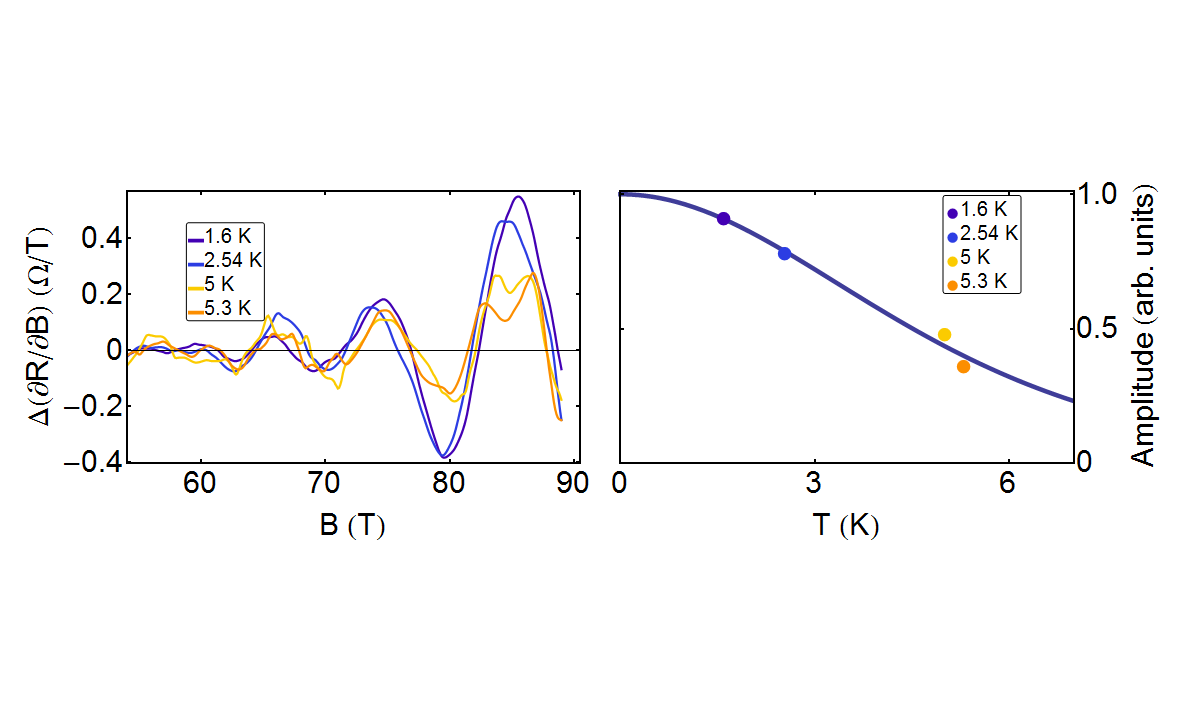}
\caption{ \textit{Left panel:} Derivative of the magnetoresistance for \YBCO{80} at each temperature with a background subtracted. \textit{Right panel:} Fit to the oscillation amplitude, yielding $\meff = 2.5\pm0.2$. }
\label{fig:4b}
\end{figure}

The derivative of the magnetoresistance for \YBCO{80} is shown in \autoref{fig:3b}, along with the background in black, plus the full fit in red. The background-free data is shown in \autoref{fig:4b}, along with the mass fit which gives $\meff = 2.5\pm 0.2$: consistent with $\mstar = 2.4 \pm 0.2$ obtained using the un-differentiated data. 

\subsection{\YBCO{86}}

The same magnetotransport technique that was performed on \YBCO{80} was also performed on \YBCO{86} (during a different experiment run, hence the different temperatures). 

\autoref{fig:7} shows the raw magnetoresistance data for \YBCO{86}, along with a fit in red and the background in black.  The fits yield $F=599 \pm 12$~T, $\Delta F = 21 \pm 10$~T, $\tau=0.07 \pm 0.03$~ps, and $\mstar = 3.6 \pm 0.2$~m$_e$. To emphasise that the background at each temperature is non-oscillatory, the derivative of the background is plotted in the right panel of \autoref{fig:6}.

The derivative of the magnetoresistance for \YBCO{86} is shown in \autoref{fig:8}, along with the background in black, plus the full fit in red. Here, the oscillations can clearly be seen in the derivative at all temperatures. The background-free data is shown in \autoref{fig:9}, along with the mass fit which gives $\meff = 3.4 \pm 0.2$: consistent with the mass of $\mstar = 3.6 \pm 0.2$ obtained using the un-differentiated data. 

\begin{figure}[H]
\includegraphics[width=\columnwidth]{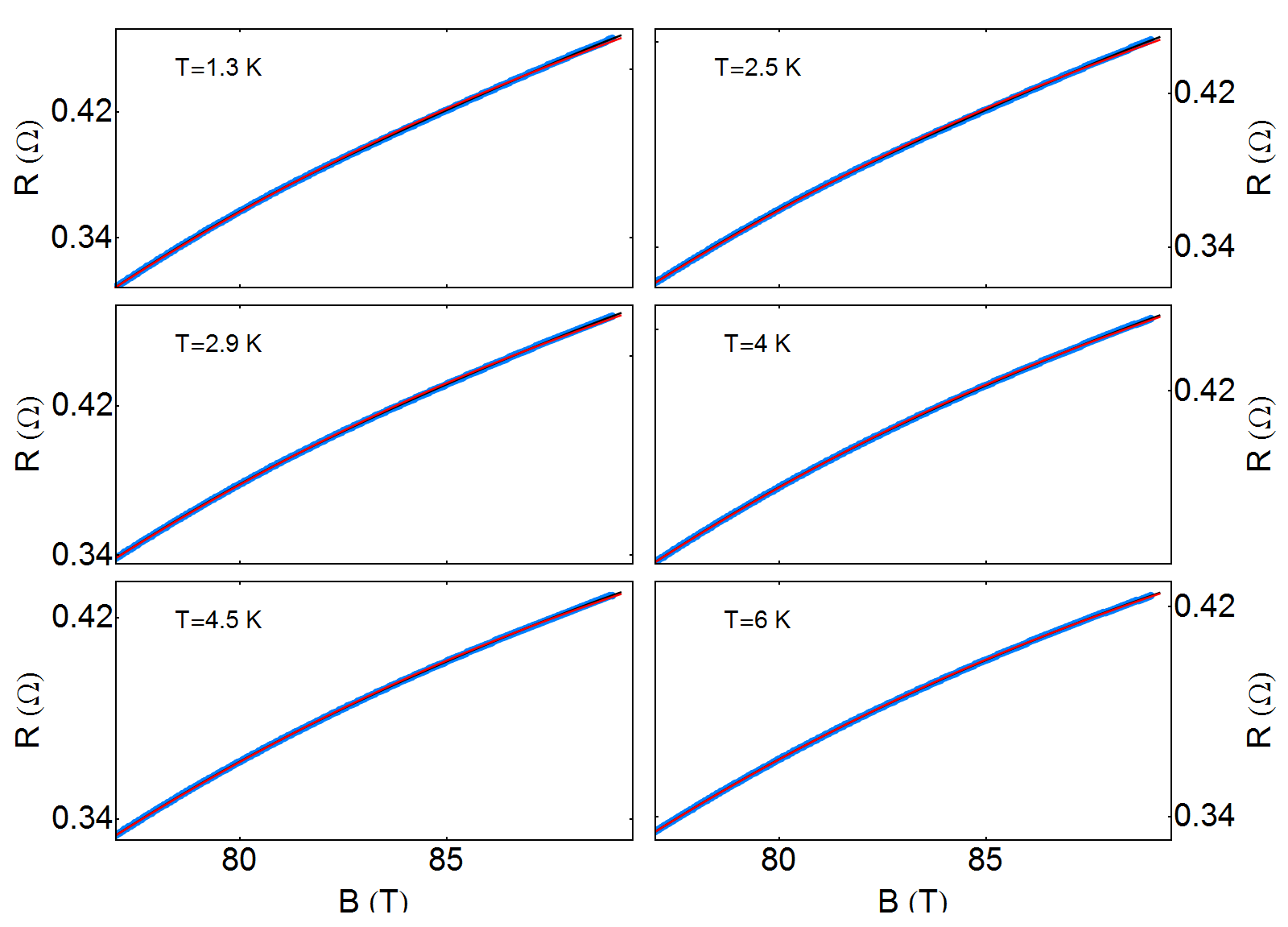}
\caption{ Magnetoresistance for \YBCO{86}: the raw data is in blue; the fit to \autoref{eq:1} plus a background is in red; the background alone is in black. All panels have the same vertical span. }
\label{fig:5}
\end{figure}

\begin{figure}[H]
\includegraphics[width=\columnwidth, clip=true, trim = 0mm 60mm 0mm 60mm]{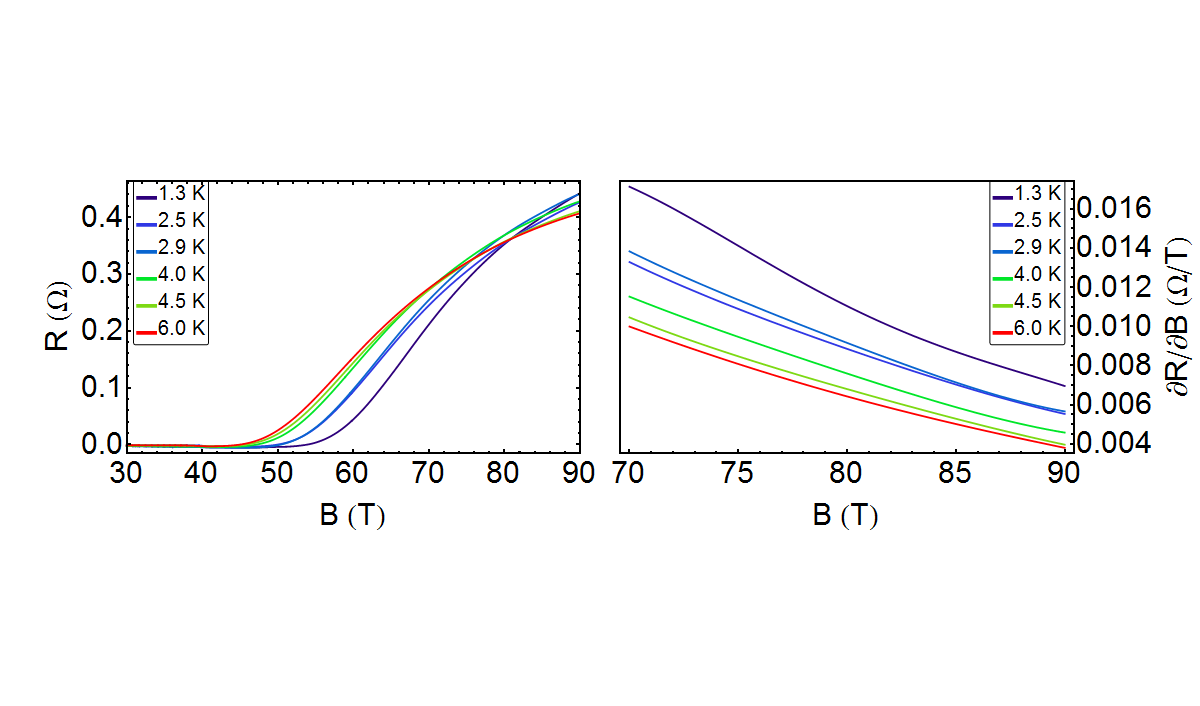}
\caption{ \textit{Left panel:} raw magnetoresistance for \YBCO{86} across the entire field and temperature range. \textit{Right panel:} derivative of the background, shown in black in \autoref{fig:5}, used to extract the oscillatory component of the magnetoresistance, show in the left panel of \autoref{fig:7}. }
\label{fig:6}
\end{figure}

\begin{figure}[H]
\includegraphics[width=\columnwidth, clip=true, trim = 0mm 60mm 0mm 60mm]{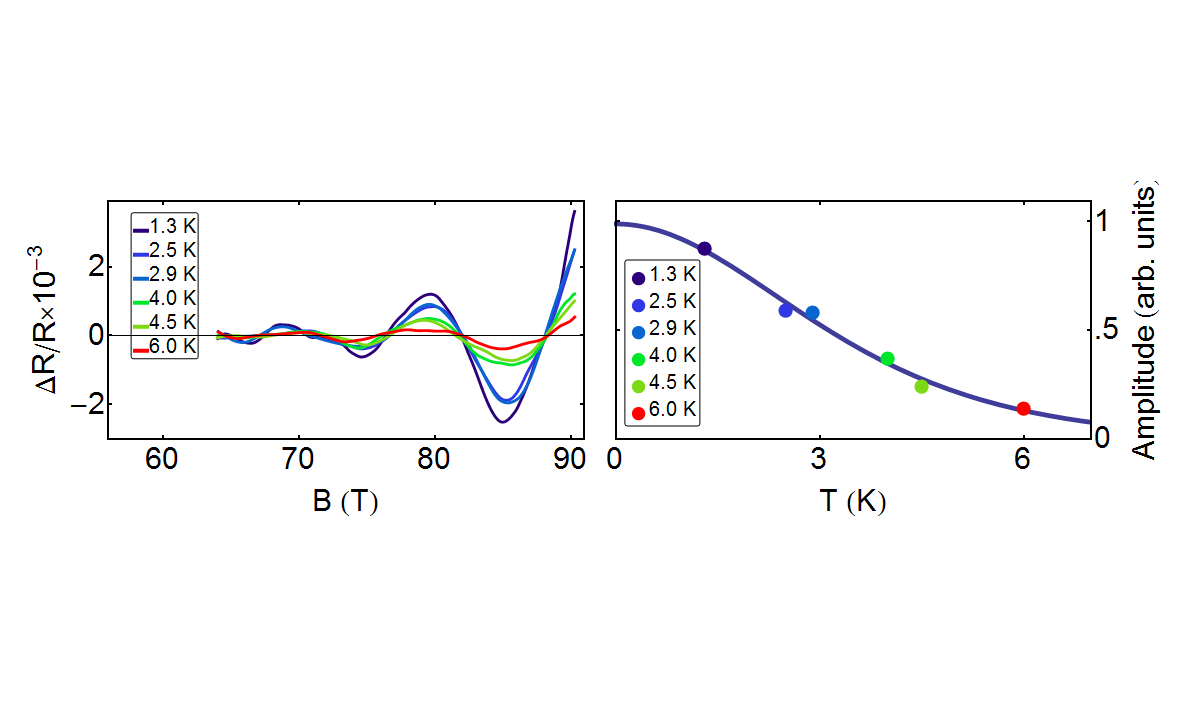}
\caption{ \textit{Left panel:} the magnetoresistance for \YBCO{86} at each temperature with a background subtracted (see \autoref{fig:7}). \textit{Right panel:} Fit to the oscillation amplitude, yielding $\meff = 3.6\pm0.2$. }
\label{fig:7}
\end{figure}

\begin{figure}[H]
\includegraphics[width=\columnwidth]{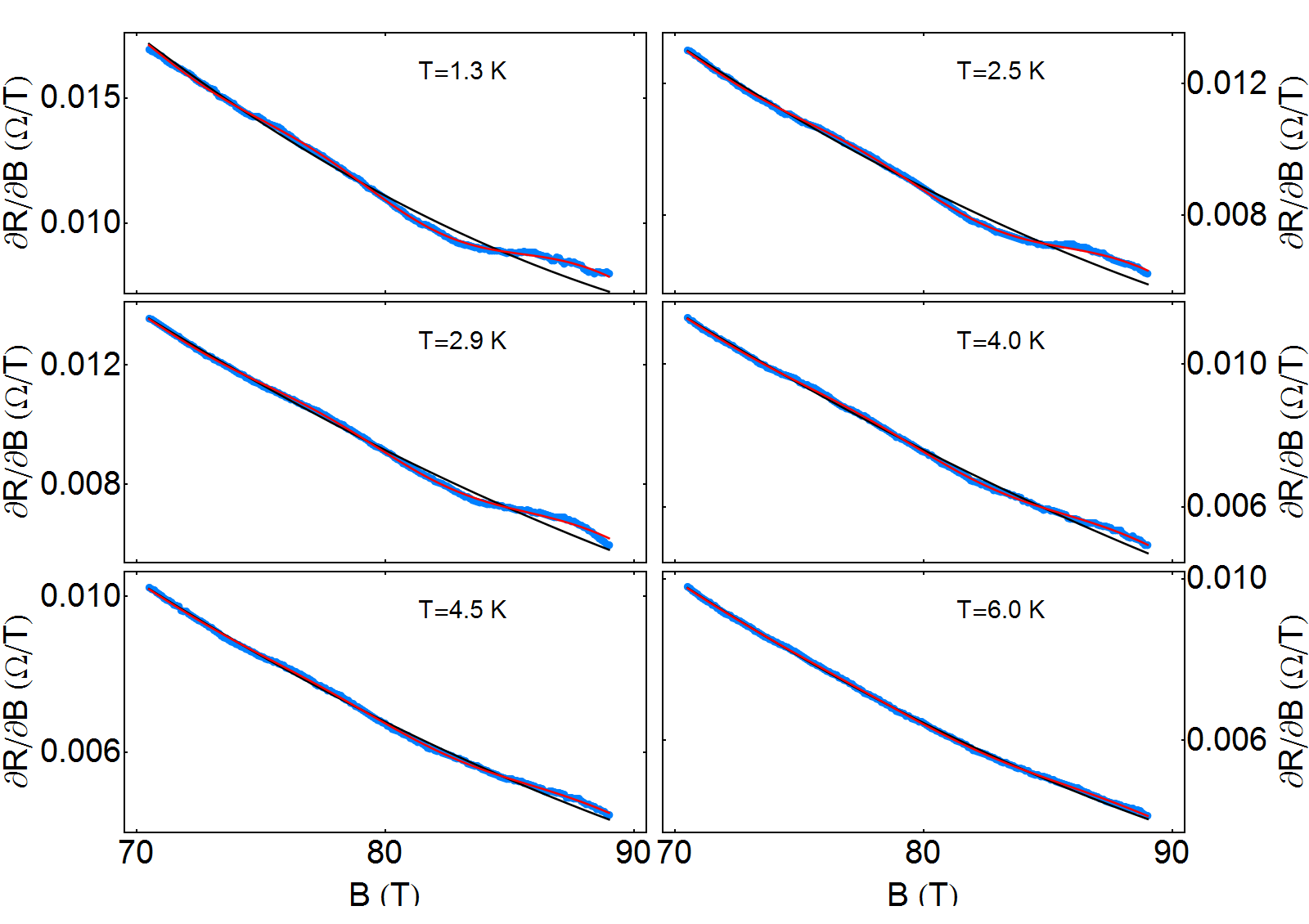}
\caption{Derivative of the magnetoresistance for \YBCO{86}: the raw data is in blue; the fit to \autoref{eq:1} plus a background is in red; the background alone is in black. All panels have the same vertical span.}
\label{fig:8}
\end{figure}

\begin{figure}[H]
\includegraphics[width=\columnwidth,clip=true, trim = 0mm 60mm 0mm 60mm]{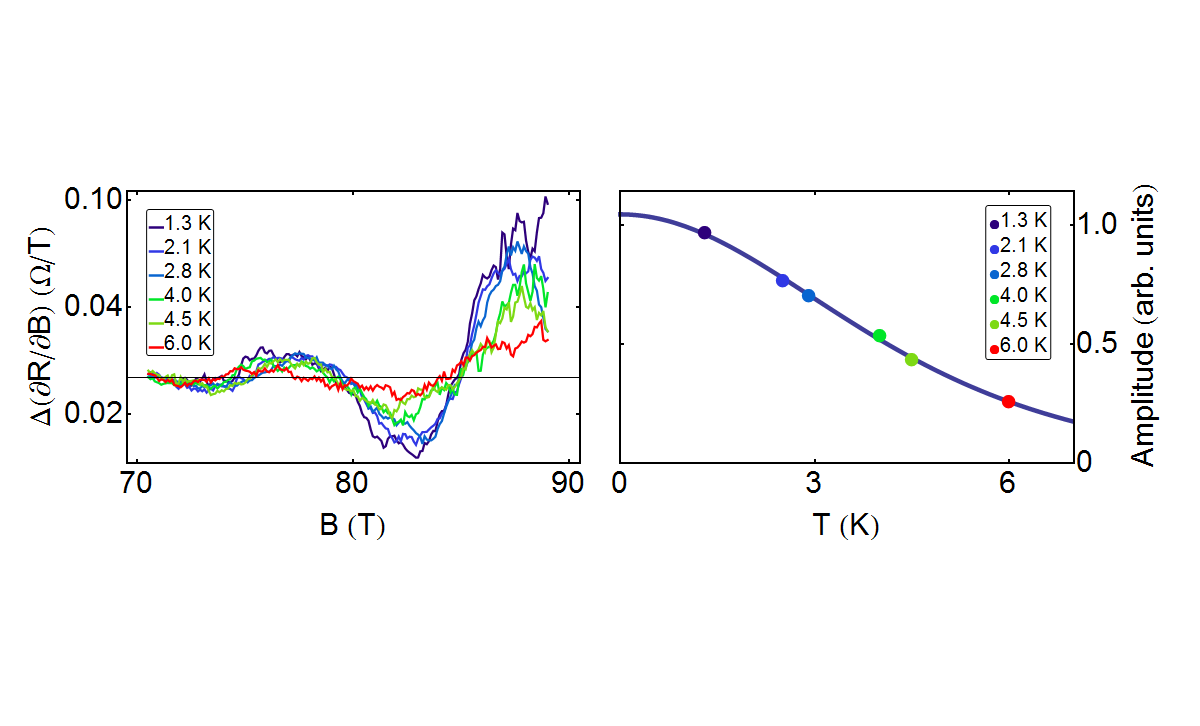}
\caption{ \textit{Left panel:} derivative of the magnetoresistance for \YBCO{86} at each temperature with a background subtracted (see \autoref{fig:3}). \textit{Right panel:} Fit to the oscillation amplitude, yielding $\meff = 3.4\pm0.2$, in agreement with the mass reported in the main text. }
\label{fig:9}
\end{figure}

\subsection{Rising and falling field}

\begin{figure}[H]
\begin{center}
\includegraphics[width=.7\columnwidth]{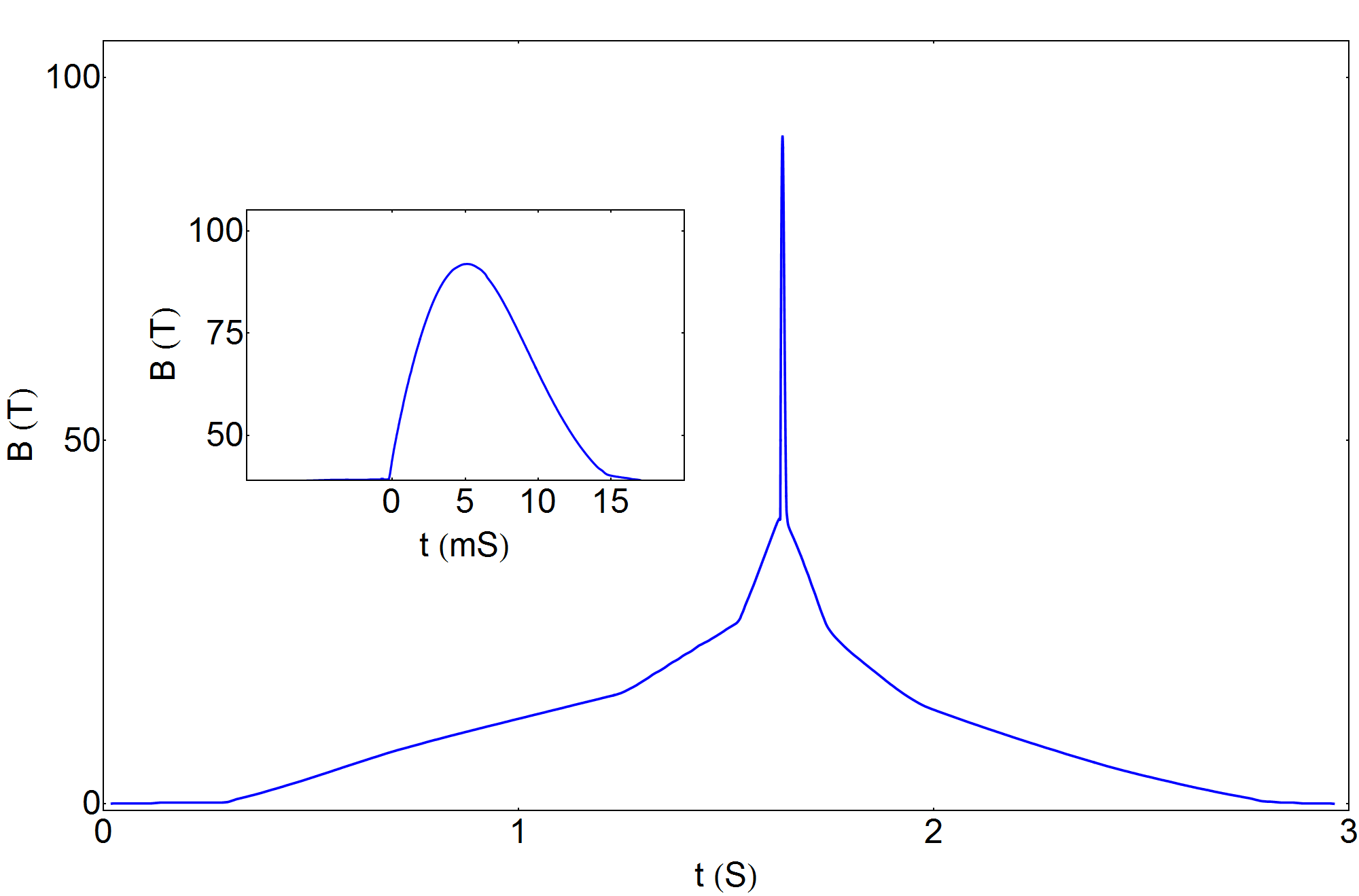}
\caption{ Field profile of the pulsed magnet used in this experiment. The ``outsert'' generator-driven magnet is ramped with a shaped waveform over 1.5 seconds to 40 T, at which point the ``insert'' capacitor-bank driven magnet is fired to 52 T, combining for 92 T at the center of the bore of the magnet.  }
\label{fig:10}
\end{center}
\end{figure}

In a pulsed field environment where $\partial B / \partial t$ exceeds $10^4$ T/s (see \autoref{fig:10}), sample self-heating can become an issue. For \highTc superconductors, the melting of the vortex lattice as the sample enters the resistive state dissipates heat in the sample. While this heating is unavoidable, the sample can usually equilibrate with the high-thermal-conductivity sapphire sample platform by peak field, and the data on falling field is then taken at a constant and known temperature. This condition is checked by looking for hysteresis at high field: if the sample has come to equilibrium by peak field, it will show no hysteresis here. This is particularly important above 4 K when the sample is in gas, rather than liquid.  This procedure has been validated by obtaining the same mass at the same doping (\YBCO{58}) in magnets of different pulse length (including the DC hybrid in Tallahassee). 

\begin{figure}[H]
\includegraphics[width=\columnwidth]{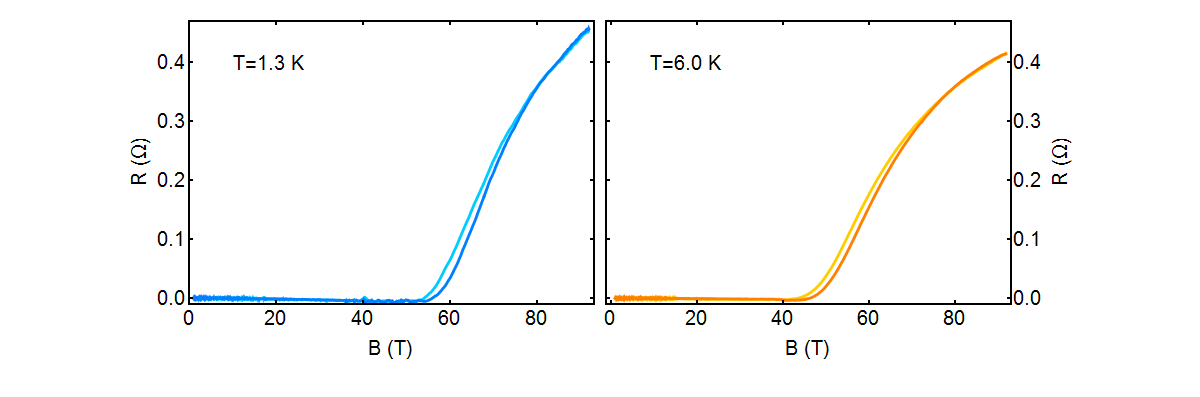}
\caption{ Rising (light shades) and falling (dark shades) field resistance for \YBCO{86} at the highest (6 K) and lowest (1.3 K) temperatures. The data here was analysed with time constant of $2 \mu s$ to avoid broadening the resistive transition on rising field. This is shorter than the $300 \mu s$ used for \autoref{fig:5} and \autoref{fig:8}, where only falling field data is shown. }
\label{fig:11}
\end{figure}

\autoref{fig:11} shows the two extremes of temperatures for \YBCO{86}: the doping where the resistive transition occurs at the highest field and equilibration time is therefore shortest. There is some heating when the resistive state is entered at both temperatures (rising data is in lighter shades), but by $\sim$75 T the hysteresis is gone, indicating thermal equilibrium. Also note that the heating is the same in the liquid (1.3 K) as it is in the gas (6 K). The size of the hysteresis at the melting transition in \autoref{fig:11}, before the sample reaches thermal equilibrium, corresponds to heating of about 1 K (see \citep{Ramshaw:2012b}).  We are therefore in the regime where the sample comes to equilibrium before peak field, and thus sample self-heating during the pulse is not a significant source of error in this experiment.  

\subsection{Physical properties at a quantum critical point}

An enhancement of the thermodynamic effective mass \meff at a quantum critical point is often accompanied by an enhancement of other physical properties that depend on the mass (density of states), including, but not limited to, the upper critical field, the superconducting condensation energy, the London penetration depth, the residual specific heat ``$\gamma$'', the ``$A$'' coefficient of the $T^2$ term in the resistivity, and the residual resistivity ``$\rho_0$'' at $T=0$ \cite{Graf:1997,Fisher:2002,Nakashima:2004,Settai:2007,Knebel:2008,Walmsley:2013}. Not all techniques show the same mass enhancement within a given material: the thermodynamic mass as measured by quantum oscillations in CeRh$_2$Si$_2$, and the specific heat $\gamma$, are enhanced by a factor of four approaching the critical pressure \cite{Graf:1997,Settai:2007}, while the square root of the $A$ coefficient ($A \propto (\meff)^2$) is enhanced by a factor of 3 over the same pressure range \cite{Settai:2007}; the thermodynamic mass in CeRhIn$_5$ is enhanced by a factor of 10 \cite{Shishido:2005}, the specific heat $\gamma$ by a factor of $\sim 6$ \cite{Fisher:2002}, the square root of \HcTwo ($\HcTwo \propto (\meff)^2$) by a factor of $\approx 2$ \cite{Knebel:2008}, and the mass estimated from the slope of \HcTwo as a function of \Tc (as tuned with pressure) by less than a factor of two \cite{Knebel:2008}; the specific heat and penetration depth in BaFe$_2$(As$_{1-x}$P$_x$)$_2$ show the same mass enhancement at the quantum critical point, but differ by a factor of two at $x \approx 0.4$ \cite{Walmsley:2013}. These differences in the apparent mass enhancement are not unexpected: each physical property is sensitive to different renormalizations. For example, the the mass measured in cyclotron resonance experiments is not enhanced by electron-electron interactions \cite{Kohn:1961}, the spin susceptibility is not enhanced by electron-phonon interactions \cite{Engelsberg:1970}, and penetration depth probes the ``dynamic effective mass'', $\frac{\meff}{1+\frac{1}{3}F_1}$ \cite{Leggett:1968}. For a review of the many different ``masses'' in metals see \cite{Leggett:1968}, section IV. Additionally the theories used to extract mass may not valid near a quantum critical point (or at any pressure/doping in an unconventional metal), for example, the BCS expression relating \HcTwo and \meff, $H_{c2} = \frac{\Phi_0}{2 \pi} \left(\frac{\pi \Delta(0) \meff}{\hbar^2 k_F}\right)^2$, will undoubtedly fail near a QCP in a non-BCS superconductor (and may only be qualitatively correct even away from the QCP).

 It must be emphasized that none of these properties truly ``diverges'' at the QCP, but instead reaches a maximum there. There are several possible explanations for this: suppression and broadening of the enhancement at the QCP due to sample inhomogeneity; the breakdown of the theories used to extract mass near the QCP, as mentioned above; the introduction of an energy scale into the system, such as a superconducting gap, that serves to cut off the fluctuating mode that is renormalizing the mass.

In the case of \YBCOd, both \HcTwo \cite{Grissonnanche:2014} and the heat capacity jump at \Tc ($\Delta \gamma$) \cite{Loram:1993,Loram:1998} show maxima near $p_{crit} \approx 0.18$ and $p_{crit} \approx 0.08$, as shown in Figure 1 of the main text. Both of these observations are consistent with the mass being enhanced at the quantum critical point at $p_{crit} \approx 0.18$. Superfluid density, as measured by $\mu$SR, on the other hand, does not show any maxima \cite{Sonier:2007} (previous measurements that do show a peak in superfluid density near $p_{crit} \approx 0.18$ have been disputed due to the introduction of zinc into these samples, which is known to be a strong breaker of Cooper pairs \cite{Bonn:1994}).  A systematic doping dependence of the normal-state $\gamma$ in \YBCOd has not been performed, due to the extreme magnetic fields needed to access the normal state.  The differences in mass enhancement seen by different probes in the cuprates may contain valuable information regarding how interactions dress the quasiparticles, and warrants further investigation.

\end{document}

%% file: macros.tex
\newcommand{\caxis}{\ensuremath{\hat{c}}-axis\xspace}

\newcommand{\wc}{\ensuremath{\omega_c}\xspace}

\newcommand{\meff}{\ensuremath{m^{\star}}\xspace}

\newcommand{\Tc}{\ensuremath{T_{c}}\xspace}
\newcommand{\highTc}{high-\ensuremath{T_c}\xspace}

\newcommand{\mstar}{\ensuremath{m^{\star}}\xspace}

\newcommand{\HcTwo}{\ensuremath{H_{c2}}\xspace}

